\documentclass[pra,twocolumn,amsmath,amssymb,floatfix]{revtex4-1}
\usepackage{graphicx}
\usepackage{epsfig}
\usepackage{dcolumn}
\usepackage{bm}
\usepackage{color}
\usepackage{titlesec}

\def\(({\left(}
\def\)){\right)}
\def\[[{\left[}
\def\]]{\right]}

\newcommand{\beq}{\begin{equation}}
\newcommand{\eeq}{\end{equation}}
\newcommand{\barr}{\begin{eqnarray}}
\newcommand{\earr}{\end{eqnarray}}
\newcommand{\bei}{\begin{itemize}}
\newcommand{\eei}{\end{itemize}}

\begin{document}

\title{Non-equilibrium dynamics of coupled Luttinger liquids}

\author{L.~Foini} 
\affiliation{Department of Quantum Matter Physics, University of Geneva, 24 Quai Ernest-Ansermet, CH-1211 Geneva, Switzerland}
\author{T.~Giamarchi} 
\affiliation{Department of Quantum Matter Physics, University of Geneva, 24 Quai Ernest-Ansermet, CH-1211 Geneva, Switzerland}

\date{\today}

\begin{abstract}
In this work we consider the dynamics of two tunnel coupled chains after 
a quench in the tunneling strength is performed and the two systems are 
let evolve independently. We describe the form of the initial state
comparing with previous results concerning the dynamics after the splitting
of a one-dimensional gas of bosons into two phase coherent systems.
We compute different correlation functions, among which those that are relevant for  
interference measurements, and discuss the emergence of 
effective temperatures also in connection with previous works.

\end{abstract}

\maketitle

\section{Introduction}

The study of the non-equilibrium dynamics in closed quantum systems
is a theoretical challenge that has been posed long time ago~\cite{VN1929,D1991,S1994,PSSV2011}
 and it is now receiving enormous 
interest thanks to the possibility of exploring this regime in the laboratories,
in particular in cold atomic experiments~\cite{BDZ2008}. 
These experiments allow one to access
the intrinsic quantum many-body dynamics of the system without dissipation and in a controllable manner.
Indeed, in this set up the system can be driven out-of-equilibrium by a sudden change of the
interaction strength among the atoms or of the confining potential.
Among others, one of the central questions in the field concerns the understanding
of the relaxation of the system and the possibility to describe the final
state as a thermal one with a well-defined effective temperature~\cite{PSSV2011,CR2010}.
 
 In this respect, thanks to their analytical tractability, one dimensional systems 
 have been discussed quite 
 extensively and have shown, in particular for integrable models, to lack equilibration
 to a Gibbs ensemble~\cite{KWW2006,RDYO2007,CEF2011,CK2012,G2013}.
 In many cases in fact correlation functions long after a quench in integrable models
 have been found compatible with a generalized Gibbs ensemble where 
 many temperatures, one for each integral of motion, are introduced.
Very recently generalized Gibbs ensembles have been observed experimentally~\cite{LEGRSKRMGS2014}.
Besides the characterization of the stationary regime, the relaxation of these systems has 
triggered lot of attention both theoretically~\cite{CC2006,cheneau2012} and experimentally~\cite{cheneau2012,LGKRS13} 
and it has been shown to occur
via a light-cone dynamics that emerges thanks to the finite velocity
of propagation of quasiparticles~\cite{CC2006,cheneau2012,LGKRS13}. 
 
Remarkable setups to explore these questions of out of equilibrium phyisics have been provided by 
atom-chips experiments~\cite{reichel2010atom,HLFSS2007,Gring_Science_2012,LGKRS13,LEGRSKRMGS2014} where 
one dimensional systems can be realized.
In such experiments a one dimensional gas of bosons is split coherently into
two systems by growing a potential barrier along  the longitudinal axes~\cite{HLFSS2007,Gring_Science_2012,LGKRS13}. 

Coupled with theoretical analysis~\cite{KISD11,SMLK2013,Gring_Science_2012,LGKRS13} 
this study revealed that, depending on the splitting process, the dynamics proceeds via a prethermalized state,
which shows certain equilibrium-like correlation functions despite being a 
non-equilibrium state. In particular, it was found that despite  the integrability of the 
model under study, the system and its correlation functions
are well described by thermal ones, with an effective temperature that is independent of 
the initial temperature of the system.  

However, while the Hamiltonian dynamics after the splitting is well defined, more difficult
is the description of the initial state as prepared in the process of splitting.
In particular different splitting process give rise to correlation functions that are better 
described by a many temperature scenario~\cite{LEGRSKRMGS2014}.

In this paper we examine a related model in which a similar analysis can be done in a controlled 
manner using Luttinger liquid theory. We look at the dynamics of two chains of bosonic 
particles that are equilibrated with a large finite coupling between the two chains
and that are subsequently let evolve independently.
  This type of protocol can be performed with cold atoms, for instance in  optical lattices 
  where ladders have been realized~\cite{AALBPB2014}.

 In particular in precedent works on the splitting of the gas~\cite{BA2007,KISD11,Gring_Science_2012} 
 the initial state was taken as a Gaussian squeezed state. 
 This form of the initial state was justified by the fact of recovering certain 
 correlation functions at time $t=0$.
In this work we give an alternative and more microscopic justification to the squeezed form
and at the same time we discuss some differences found between the state 
that was used in~\cite{BA2007,KISD11,Gring_Science_2012} 
and the one that we derive in this work.

Moreover we also analyze the relaxation of the system from the time dependent to the stationary state
and we discuss the possibility to define an effective temperature long after the quench, by inspecting
several correlation functions.

The paper is organized as follows: in section~\ref{Sec_model} we introduce the
model under study and the way it is analytically treated; in section~\ref{Sec_squeezed_state}
we discuss the form of the initial state considered in previous works 
focussing on the splitting of the condensate, in section~\ref{Sec_quench}
we describe the initial state that is recovered considering a quench 
of the tunneling term between the two chains and in section~\ref{Sec_Comparison} 
we discuss the comparison between the two forms; in section~\ref{Sec_Teff}
we discuss the occupation of the modes after the quench; in section~\ref{Sec_phase}
we derive the correlation functions which are relevant for interference experiments
and in section~\ref{Sec_density} we compute correlation functions associated to
the density and the current. Finally in section~\ref{Sec_discussions} we discuss our results 
and section~\ref{Sec_conclusions} summarizes the work giving some perspectives.

\section{Interacting condensates: quadratic approximation}\label{Sec_model}

The system is prepared as two tunnel coupled one-dimensional chains. 
The Hamiltonian of the 
two systems at time $t=0$ is well described within the Luttinger liquid theory by~\cite{giamarchi2004}:
\beq
\displaystyle H^{1+2} = H_{LL}^1 + H_{LL}^2 - \, \frac{t_{\perp}}{2\pi}  \, \int {\rm d} x \, 2  \cos( \theta_2(x) - \theta_1(x)) \ .
\eeq
Here $H_{LL}^\alpha$ is the Luttinger liquid Hamiltonian describing each chain $\alpha=1,2$:
\beq
\displaystyle H_{LL}^\alpha =\frac{\hbar u}{2} \int {\rm d} x \, \Big[ \, \frac{\mathcal{K}}{\pi} \, [\nabla \theta_{\alpha}(x) ]^2 \, + \,\frac{\pi}{\mathcal{K}} [n_{\alpha}(x) ]^2 \, \Big] \ .
\eeq
$\mathcal{K} $ is the Luttinger parameter and $u$ the sound velocity. 
The effects of interactions are hidden in those two parameters. 
In the weakly interacting regime the Luttinger parameter is  $\mathcal{K}= \hbar\pi \sqrt{\frac{\rho}{m g}}$ in terms of the microscopic parameters of the gas,
where $g$ is the effective interaction constant and $\rho$ the density.  The hard core limit is instead achieved
for $\mathcal{K} = 1$. The sound velocity is given by $u=\hbar \pi \rho/ m \mathcal{K}$
and therefore $u=\sqrt{ \frac{\rho g}{m} }$ in the weakly interacting regime.
The operators $\theta_{\alpha}(x)$ and 
$n_{\alpha}(x)$ representing respectively the 
phase of the bosonic field and the fluctuation of its density in the  
system $\alpha=1,2$ 
are canonically conjugated:
$[n_{\alpha}(x),\theta_{\beta}(x')]=i \delta(x-x') \delta_{\alpha,\beta}$.
The cosine term originates from the tunneling operator $\psi_1^{\dag}(x)\psi_2(x) + \text{h.c.} = 2 \, \rho  \cos( \theta_2(x) - \theta_1(x))$,
where we used that $\psi_\alpha(x) \simeq \sqrt{\rho} \, e^{i \theta_\alpha(x)}$. Therefore one can set $t_{\perp} = \,\hbar \,J \rho$. 
 
 The dynamics of this system can be studied by introducing symmetric and antisymmetric variables 
 $\theta_{A/S} = \theta_1 \mp \theta_2$ and $n_{A/S} = (n_1 \mp n_2)/2$. 
 Indeed under the assumption that the two systems are identical, symmetric and antisymmetric modes 
  decouple and in terms of those variables one has $H = H_{LL}^S + H_{SG}^A$
with $H_{SG}^A $ the Sine-Gordon Hamiltonian for the antisymmetric modes:
\beq\label{HA}
\begin{array}{ll}
\displaystyle H_{SG}^A =  & \displaystyle \frac{\hbar u}{2} \, \int {\rm d} x \, \Big[ \, \frac{K}{\pi} [\nabla \theta_A(x)]^2 + \frac{\pi}{K} [n_A(x)]^2 \, \Big] 
\\ \vspace{-0.2cm} \\
& \displaystyle \qquad- \, \frac{t_{\perp}}{2\pi}  \, \int {\rm d} x \, 2 \, \cos \, \theta_A(x) \ ,
\end{array}
\eeq
where $K= \mathcal{K}/2$.
In the following we will focus only on the antisymmetric part of $H$.

A limit which is particularly relevant for the next discussion is the case
of $t_{\perp}=0$ when $H^A_{SG}$ reduces to the Luttinger liquid Hamiltonian
which in its diagonal form reads:
\beq\label{HLL}
\displaystyle H^A_{LL} = \sum_{p\neq 0} \hbar u |p| b^{\dag}_p b_p + \frac{\hbar u\pi}{2K} n_0^2 \ ,
\eeq
 made of non interacting sound wave modes. In order to obtain (\ref{HLL})
we have expanded the fields $n_A(x)$ and $\theta_A(x)$ over the bosonic operators that diagonalize 
$H^A_{LL}$:
\beq
\begin{array}{ll}
\displaystyle  \theta_A(x) = &  \displaystyle \frac{i}{\sqrt{L}}  \sum_{p \neq 0} \  e^{ - i p x} e^{- \alpha^2 p^2 /2}
\sqrt{\frac{\pi}{2 K |p|}} (   b^{\dag}_{p} -b_{-p}) 
\\ \vspace{-0.2cm} \\
&  \displaystyle \qquad+ \, \frac{1}{\sqrt{L}} \theta_{0} 
\end{array}
\eeq
\beq
\begin{array}{ll}
 \displaystyle n_A(x) = &  \displaystyle \, \frac{1}{\sqrt{L}}  \sum_{p\neq 0} e^{ - i p x} e^{- \alpha^2 p^2 /2}
\sqrt{\frac{|p| K}{2 \pi}} (b^{\dag}_{p} +  b_{-p} ) 
\\ \vspace{-0.2cm} \\
&  \displaystyle \qquad + \, \frac{1}{\sqrt{L}} n_{0} \ .
\end{array}
\eeq
where $\alpha$ is a cutoff regularizing the integrals. 

We treat  the cosine term in (\ref{HA}) for $t_{\perp}\neq0$ 
by making a semiclassical expansion in $\theta_A$ around its minimum $\theta_A(x)=0$.
This approximation 
greatly simplifies the problem, 
turning the Hamiltonian into a quadratic one.
Indeed, under these conditions,  the Hamiltonian reads  (up to irrelevant constants):
\beq\label{H_quadratic}
\begin{array}{ll}
 \displaystyle H^A_{SC} &= \displaystyle  \frac{u\hbar}{2} \int {\rm d} x \, \Big[ \, \frac{K}{\pi} \, [\nabla \theta_{A}(x) ]^2 \, + \,\frac{\pi}{K} [n_{A}(x) ]^2 \, \Big]  
\\ \vspace{-0.2cm} \\
&\qquad\qquad \displaystyle  + \frac{t_{\perp}}{2 \pi}  \int {\rm d} x  (\theta_A(x))^2   \ .
\end{array}
\eeq
Such semiclassical expansion is well adapted when $K$ is very large, as in the experiments~\cite{Gring_Science_2012}.
In order to diagonalize the Hamiltonian $H^A_{SC}$ (\ref{H_quadratic}) for generic $t_{\perp}$
we note that under the quadratic approximation 
all $p$-modes are decoupled. The Hamiltonian (\ref{H_quadratic})  
is made diagonal through the following transformation involving 
the bosonic modes with $p\neq0$:
\beq\label{Bogoliubov}
\begin{array}{ll}
 \displaystyle \eta_p^{\dag} = \cosh \varphi_p b_p^{\dag} - \sinh\varphi_p b_{-p}
\\ \vspace{-0.2cm} \\
 \displaystyle \eta_{-p} = \cosh \varphi_p b_{-p} - \sinh\varphi_p b_{p}^{\dag} \ .
\end{array}
\eeq
The operators $\{\eta_p,\eta^{\dag}_p\}$ satisfy canonical bosonic commutation 
relations and $\varphi_p=\varphi_{-p}$ is fixed by the condition:
\beq
  \tanh 2\varphi_p=  \frac{t_{\perp}}{t_{\perp} + 2 K \hbar u |p|^2} 
  \qquad \varphi_p = \frac12 \log \frac{\omega_p}{u |p|}
  \ ,
\eeq
with $\omega_p = \sqrt{(\hbar u p)^2 + \frac{t_{\perp} \hbar u}{K}} =  \sqrt{(\hbar u p)^2 +m^2} $,
where we set  $\displaystyle m^2 \equiv \frac{\hbar u t_{\perp}}{K}$.
The $p=0$ component of the Hamiltonian is made diagonal by the transformation:
\beq
\eta_{p=0}^{\dag} = \frac{1}{\sqrt{2}} ( \frac{n_0}{\sqrt{w}} - i \sqrt{w} \, \theta_0 )
\eeq
with $w=\frac{m K}{\hbar u \pi} $.
With these transformations the Hamiltonian becomes:
\beq\label{HASC}
\begin{array}{ll}
 \displaystyle H^A_{SC} & \displaystyle =      \sum_{p} \hbar \omega_p  \eta_p^{\dag}\eta_p  \ .
\end{array}
\eeq
\noindent
With respect to the Luttinger liquid Hamiltonian we see that 
the interaction $t_{\perp}$ in (\ref{HASC}) opens a gap  and turns the spectrum $\omega_p$ 
at small $p$ into a quadratic one. 

\section{Initial state}

\subsection{Phenomenological derivation}\label{Sec_squeezed_state}

As we mentioned in the introduction the theoretical
framework of the previous section, and in particular the dynamics at $t_{\perp}=0$,
 is relevant to be compared with the description of experiments where a single gas is split into two  gases
which are let evolve independently.
In the literature the initial state after the splitting of the system is taken of the following form~\cite{KISD11,Gring_Science_2012}:
\beq\label{psi0}
\displaystyle |\tilde{\psi}_0\rangle = \frac{1}{\mathcal N} \exp \Big[\sum_{p>0} \tilde{W}_p b_p^{\dag} b_{-p}^{\dag}\Big] |0\rangle  |\tilde{\psi}_{p=0}\rangle 
\eeq
and 
$\displaystyle  \tilde{W}_p = \frac{1 - \tilde{\alpha}_p}{1 + \tilde{\alpha}_p} $, $\displaystyle \tilde{\alpha}_p = \frac{|p| K}{\pi \rho} $, 
$\langle n_0  |\tilde{\psi}_{p=0}\rangle = e^{-\frac{1}{2\rho} n_0^2}$.
Here $K$ is the Luttinger parameter and $\rho$ the density.  
$\mathcal N$ is the normalization of the state $\mathcal N = \prod_{p>0} (1 - \tilde{W}_p^2)^{-1/2}$. 
This form of the initial state is taken to reproduce the following correlation functions:
\beq\label{Initial_correlations}
\begin{array}{ll}
\displaystyle \langle n_A(p) n_A(p')  \rangle(t=0) =  \frac{\rho}{2}  \, \delta_{-p,p'} \ ,
\\ \vspace{-0.2cm} \\
\displaystyle \langle \theta_A(p) \theta_A(p') \rangle(t=0) =  \frac{1}{2 \rho} \, \delta_{-p,p'} \ ,
\end{array}
\eeq
resulting in local correlations in real space $\langle n_A(x) n_A(x')  \rangle(t=0) =  \frac{\rho}{2}  \, \delta(x-x')$.
These correlation functions should be considered with the delta function 
smeared over the healing length scale $\xi_h = \hbar  u / g \rho$.
This form of state is chosen assuming that the splitting leads to a random process where particles can go either left or right
with equal probability. 
In the limit of large number of particles this process results in a Gaussian distribution of particle number difference
where density fluctuations after the splitting are chosen
to be proportional to the density itself. The strength of phase fluctuations follows 
considering the state as a minimum uncertainty state.
In the following we will give a different explanation of how a state of the form (\ref{psi0}) should be expected.
The analogous coefficient $\tilde{W}_p$ will be slightly different.

\subsection{Quench in $t_\perp$}\label{Sec_quench}

In this section we take as initial state the one resulting from  a sudden quench of $t_{\perp}$.
This means that we assume that the initial state is prepared equilibrating the system in the ground state
of the Hamiltonian (\ref{H_quadratic}) at large $t_{\perp}$ and then suddenly change $t_{\perp}$
of a finite amount. In particular we will compare this result with the one of the previous section 
in the case where the dynamics
is driven with $t_{\perp}=0$. Hereafter we will set $\hbar=1$. 
A similar situation concerning a quench of the mass in a bosonic field theory was considered in~\cite{SC2010,SGS2013,IC2010}.

The Hamiltonian at time $t=0$ is associated with a bosonic operator
$\eta_p^{0}$ such that the initial state is the vacuum of this operator $\eta_p^{0} |\psi_0\rangle=0$.

One can show that in terms of the bosonic operators $b_p^{\dag}$, $b_p$ diagonalizing the Hamiltonian
(\ref{H_quadratic}) with $t_{\perp}=0$ the initial state has a squeezed form of the type (\ref{psi0}):
\beq\label{psi00}
\displaystyle |\psi_0\rangle = \frac{1}{\mathcal N} \exp \Big[\sum_{p>0} W_p b_p^{\dag} b_{-p}^{\dag}\Big] |0\rangle  |\psi_{p=0}\rangle \ ,
\eeq
where the coefficients $W_p$ are now different from (\ref{psi0}).

Indeed, one can compute the action of the operator $\eta_p^0$ over a state with a squeezed form as in (\ref{psi00}), 
after performing a Bogoliubov transformation as in (\ref{Bogoliubov}).
Taking $|\psi_0\rangle$ as in (\ref{psi00}) and considering $p\neq 0$ one obtains:
\beq
\begin{array}{ll}
 \displaystyle\eta_p^0 |\psi_0\rangle &  \displaystyle =  \Big( \cosh \varphi_p^0 b_p - \sinh\varphi_p^0 b_{-p}^{\dag} \Big)  |\psi_0\rangle 
\\ \vspace{-0.2cm} \\
&  \displaystyle = \cosh\varphi_p^0 W_p b_{-p}^{\dag} |\psi_0\rangle - \sinh \varphi_p^0 b_{-p}^{\dag} |\psi_0\rangle \ .
\end{array}
\eeq
Imposing $ \displaystyle\eta_p^0 |\psi_0\rangle = 0$ one finds $W_p = \text{tanh} \,\varphi_p^0 = \frac{1 - \alpha_p}{1+\alpha_p}= 
\frac{1 - \frac{u |p|}{\omega_p^0}}{1+ \frac{u |p|}{\omega_p^0}}$. This form of squeezed state was also found in~\cite{SGS2013}.
The component $p=0$ can be written as $|\psi_{p=0}\rangle\propto e^{- \frac{n_0^2}{2 w}}|0\rangle$ with $\theta_0|0\rangle = 0$.
In fact
\beq
\eta_{p=0}^0|\psi_{p=0}\rangle = \frac{1}{\sqrt{2}} ( \frac{n_0}{\sqrt{w}} + i \sqrt{w} \, \theta_0 ) |\psi_{p=0}\rangle = 0
\eeq
where we used $[\theta_0, n_0]=- i$ and $[\theta_0, f(n_0)] = [\theta_0,n_0] f'(n_0)$.

\subsection{Comparison of the two forms}\label{Sec_Comparison}

Let us now compare the results of the phenomenological derivation (\ref{psi0}) with our ladder case (\ref{psi00}). 
Considering the values given after (\ref{psi0}) one has to compare $\alpha_p = \frac{u |p|}{\omega_p^0}$ with
$\tilde{\alpha}_p = \frac{|p| K}{\pi \rho}$. 

The two expressions present points of similarity and of difference. Indeed, to go from one to the other
one has to change $\omega_p^0 \to \frac{\pi \rho u}{K}$. In the limit of non-zero tunneling (which is the case for our 
initial state) and small momenta one can approximate $\omega_p^0 \simeq m^0 = \sqrt{\frac{J_0 \rho u}{K}}$

\noindent
This calculation shows that a quench in $t_{\perp}$ would lead to a state with a squeezed form in $b_p^{\dag}b_{-p}^{\dag}$ and the coefficients 
that agree with the one used in~(\ref{psi0}) in the limit of small momenta for what concerns the dependence in $p$
but not in the coefficient of proportionality. 

\section{Effective temperature}\label{Sec_Teff}

It has been shown that the state (\ref{psi0}) leads to correlation functions
that are well described by an effective temperature $T_{\rm eff}=\frac{\pi u \rho}{4 K}$~\cite{SMLK2013,Gring_Science_2012}.

\begin{figure}
  \begin{center}
        \includegraphics[height=2.5cm]{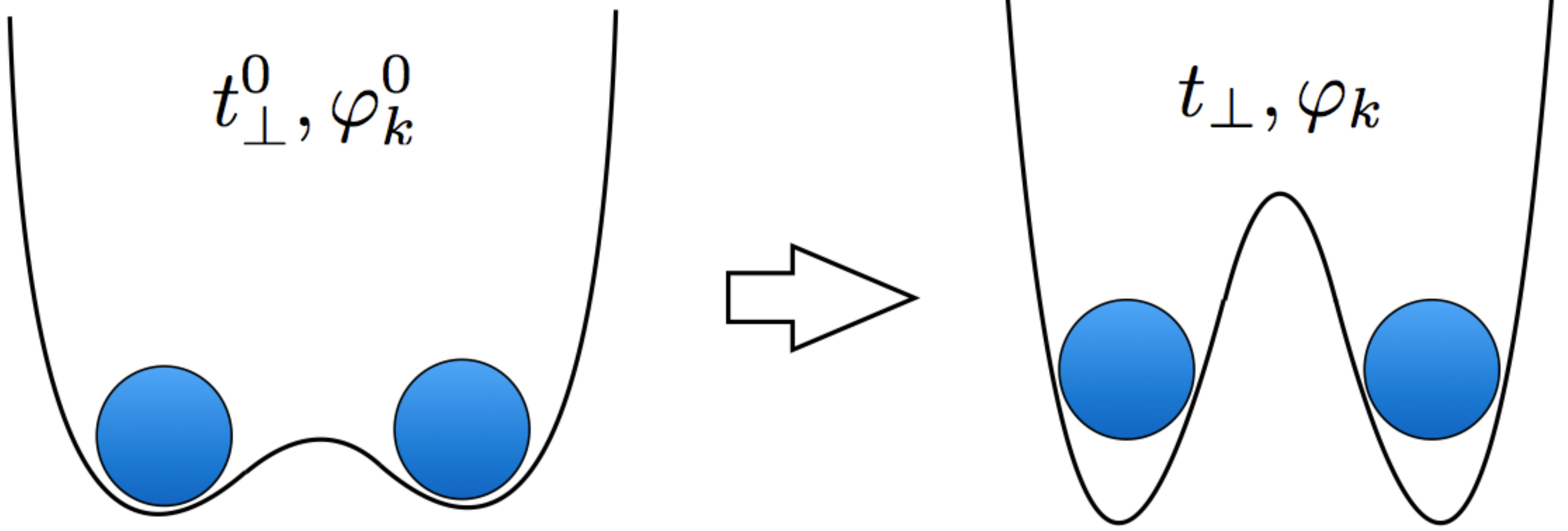}
                 \caption{ Schematic figure representing the quench from high to low tunnel-coupled systems.
    The two states are    characterized respectively  by a given tunneling strength $t_{\perp}^0$ and $t_{\perp}$ and Bogolioubov angles $\varphi^0_k$
                and $\varphi_k$. 
        }\label{FIG0}
  \end{center}
\end{figure}

In order to see the differences in the effective temperature that emerges between $|\tilde{\psi}_0\rangle$ in (\ref{psi0}) and 
$|\psi_0\rangle$ in (\ref{psi00}) 
we now consider a quench from the squeezed state characterized by  $\varphi_p^0$ 
towards a situation where the barrier is higher and the bosonic operator are associated
with an angle $\varphi_p$ (see Fig,~\ref{FIG0}). The case of independent systems and thus of infinite barrier 
corresponds to $\varphi_p=0$. 
The mapping between the operators diagonalizing the pre and post quench Hamiltonian 
reads as follows: 
\beq
\begin{array}{ll}
 \displaystyle\eta_p & \displaystyle   
= \cosh \Big(  \varphi_p   -  \varphi_p^0 \Big) \, \eta_{p}^0   - 
\sinh \Big(  \varphi_p   -  \varphi_p^0 \Big)\, \eta_{-p}^{0\dag} 
\end{array}
\eeq
We note that: 
\beq
 \displaystyle \frac{\Delta_p}{2} = \varphi_p   -  \varphi_p^0 = \frac12 \log \frac{\omega_p }{\omega_p^0} 
\eeq
and the occupation of the modes reads:
\beq
\begin{array}{ll}
 \displaystyle\text{cotanh} \frac{\beta_{\rm eff}(p) \omega_p}{2} &  \displaystyle= 2 \langle \eta_p^{\dag}\eta_p\rangle + 1  = \frac{1}{2} \Big(  \frac{\omega_p}{\omega_p^0} \, 
+  \frac{\omega_p^0}{\omega_p}\, \Big) 
\\ \vspace{-0.2cm} \\
&  \displaystyle = \cosh \Delta_p \ ,
\end{array}
\eeq
which is the analogue of the formulas derived within other quenches and quadratic
models \cite{CEF2011,FCG2011}.
This leads to $p$-dependent effective temperatures:
\beq\label{betaGGE}
\beta_{\rm eff}(p) = \frac{1}{\omega_p} \, \log \, \frac{\cosh\Delta_p+1}{\cosh\Delta_p-1}
 = \frac{2}{\omega_p} \, \log \, \frac{\omega_p+\omega_p^0}{|\omega_p-\omega_p^0|}
\eeq
as found in \cite{SCC09}. 
If one takes the limit of small momenta and assume that $m^0 \gg m$ this formula gives 
$T_{\rm eff} = \frac{m^0}{4}  = \frac14\sqrt{\frac{J_0 u \rho }{K}}$.
One should therefore compare this temperature with the one found in \cite{KISD11,Gring_Science_2012}  
which is $\tilde{T}_{\rm eff} = \frac{\pi u \rho}{4 K}$. Clearly, the two formulas 
show different dependences on the density, the sound velocity and the Luttinger parameter.

\section{Dynamics of the phase}\label{Sec_phase}

In the following we focus on the dynamics after the quench of $t_{\perp}$ and 
 consider correlation functions that are relevant for interference experiments.
We define $n_p= \sqrt{\frac{|p| K }{2\pi}} (b^{\dag}_p+b_{-p})$ and $\theta_p = i \sqrt{\frac{\pi}{2K |p|}} (b^{\dag}_p - b_{-p})$.
Under the quench protocol described in the previous section correlation functions at time $t=0$ have the following form:
\beq\label{EqAverages}
\begin{array}{ll}
\displaystyle \langle\psi_0|  n_p n_{p'} |\psi_0\rangle(t=0) = \delta_{p',-p}  \frac{K}{2 \pi u} \omega_p^0
\\ \vspace{-0.2cm} \\
\displaystyle \ \langle\psi_0| \theta_p \theta_{p'}|\psi_0\rangle (t=0) =  \delta_{p',-p}  \frac{\pi u}{2 K} \frac{1}{\omega^0_p} \ .
\end{array}
\eeq
We consider the dynamics of the system when $t_{\perp}=0$ and the two condensates evolve independently.

\begin{figure}
  \begin{center}
        \includegraphics[height=6cm]{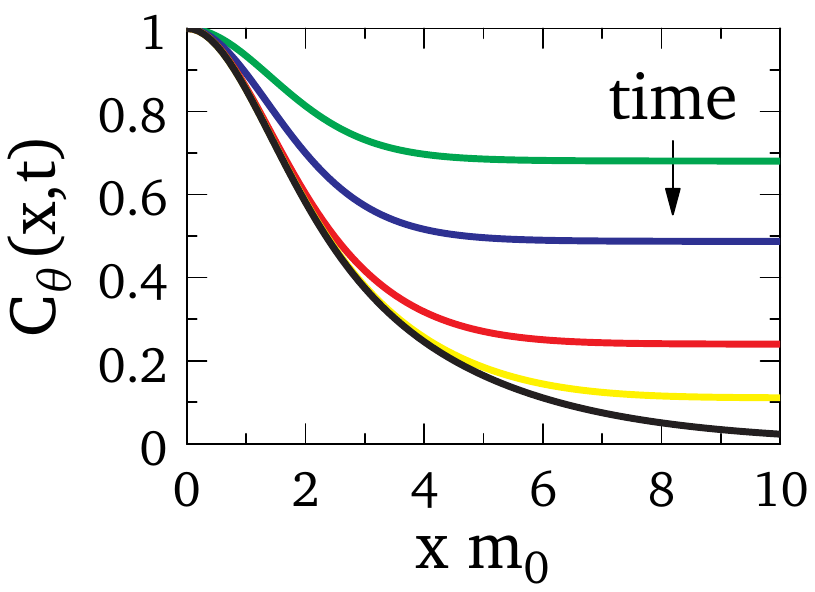}
            \includegraphics[height=6cm]{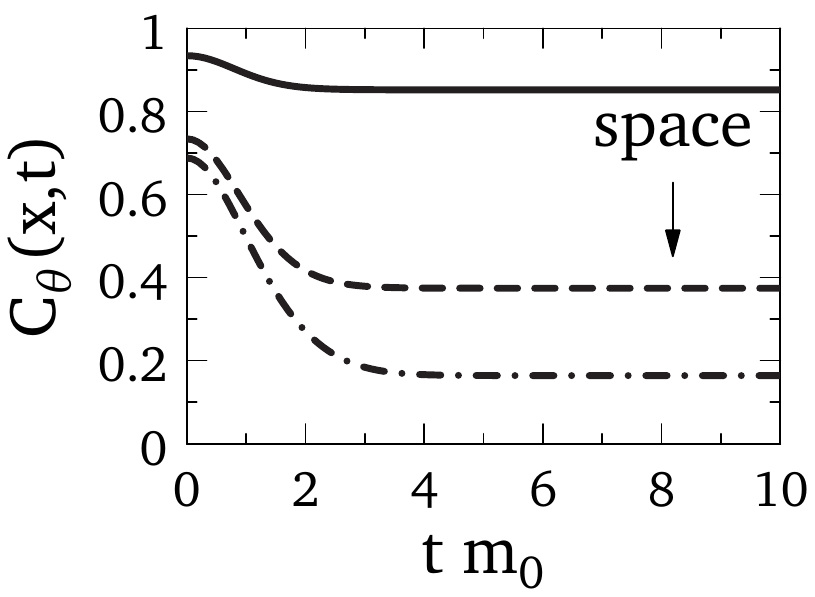}
        \caption{Correlation functions $C_{\theta}(x,t)$~(\ref{CA}),
for different time and length scales. Here $K=1$, $u=1$ and $\alpha m_0=1$. 
Upper panel: $C_{\theta}(x,t)$ as a function of the space distance for four
different times $t m_0=0.1,1,2,3$ shown respectively with green, blue, red and yellow lines. The black envelope
is the asymptotic stationary correlation function.
Lower panel: $C_{\theta}(x,t)$ as a function of the time, for three different space distances $x m_0=1, 3,5$ shown 
respectively with solid, dashed and dotted-dashed lines.
        }\label{FIG1}
  \end{center}
\end{figure}

From the interference one is able to extract the following  correlation functions~\cite{PAD2006,Gring_Science_2012,LGKRS13}:
\beq
C_{\theta}(x,t) = \frac{\langle \psi_1(x)\psi^{\dag}_2(x)\psi^{\dag}_1(0)\psi_2(0)\rangle}{\langle |\psi_1(x)|^2\rangle \langle |\psi_2(x)|^2\rangle }
\eeq
which involves the relative phase between the two condensate and reads:
\beq\label{CA}
C_{\theta}(x,t) = \langle e^{i [\theta_A(x,t) - \theta_A(0,t) ]}\rangle = e^{ - \frac12 \langle [\theta_A(x,t) - \theta_A(0,t) ]^2\rangle} 
\eeq
where:
\beq\label{EqTheta}
\begin{array}{ll}
 \displaystyle 
\langle [\theta_A(x,t) - \theta_A(0,t) ]^2\rangle 
  \displaystyle  =  \frac{1}{\pi}   \int  {\rm d}p \, (1 - \cos px)  e^{- \alpha^2 p^2} \times
\\ \vspace{-0.2cm} \\
  \displaystyle  \times  \Big[ \sin^2( u |p| t) \,  \frac{\pi^2}{K^2  p^2} \, \langle |n_p|^2 \rangle (0) + \,  \cos^2(u |p| t) \, \langle |\theta_p|^2 \rangle (0)  \Big] \ ,
\end{array}
\eeq
and $\alpha$ is a small cutoff regularizing the integral. 
In the upper panel of Fig.~\ref{FIG1} we show the correlation function (\ref{CA}) as a function of
space for four different times and the stationary limit. The lower panel shows the same correlation
function as a function of time for three 
different space differences.

\begin{figure}
  \begin{center}
        \includegraphics[height=6cm]{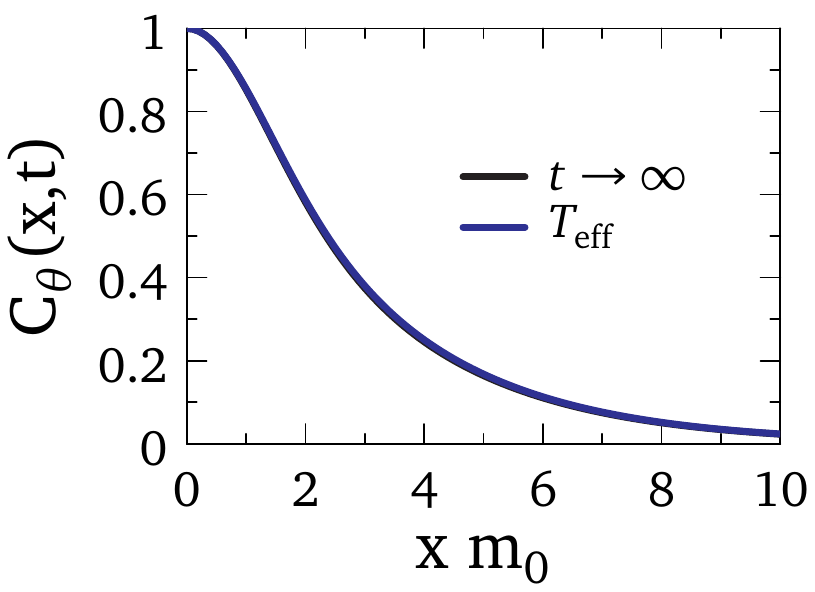}
    \includegraphics[height=6cm]{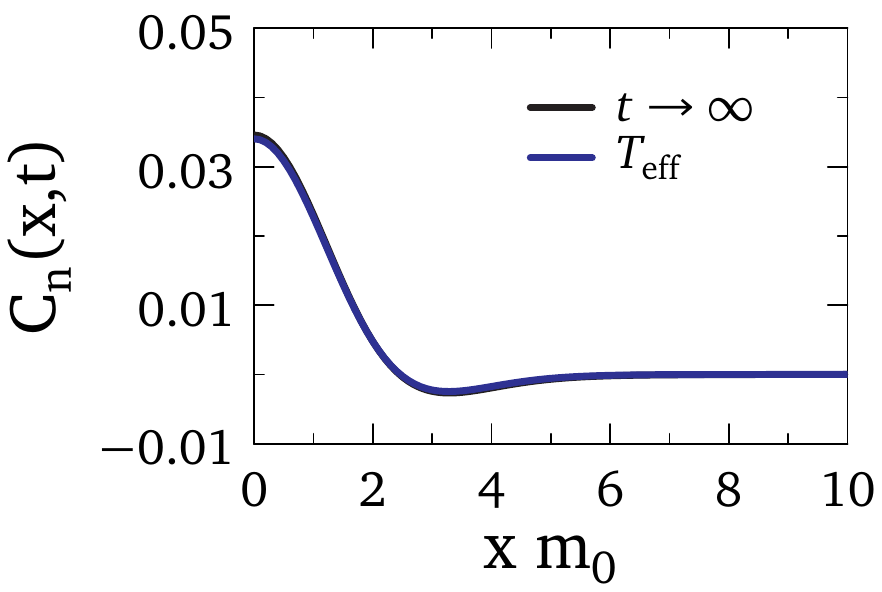}
        \caption{Upper panel: asymptotic correlation functions $C_{\theta}(x,t\to\infty) = \exp(- \frac12 \lim_{t\to\infty} \langle [\theta_A(x,t) - \theta_A(0,t) ]^2\rangle) $
        with $\lim_{t\to\infty} \langle [\theta_A(x,t) - \theta_A(0,t) ]^2\rangle$ given in 
        ~(\ref{CAsymptotic}) and thermal correlation function 
        $C_{\theta}(x) = \exp(- \frac12 \langle [\theta_A(x) - \theta_A(0) ]^2\rangle_{T_{\rm eff}} )$ with $\langle [\theta_A(x) - \theta_A(0) ]^2\rangle_{T_{\rm eff}}$ given 
        in (\ref{CThermal})
shown with black and blue lines respectively. 
Lower panel: asymptotic correlation function of the relative density~(\ref{Cnxtinfty}) (black line) compared with the equilibrium one~(\ref{Cneq}) (blue line).
In both plots $\alpha m_0 =1$, $K=1$, $u=1$ and $T_{\rm eff} = \frac{m_0}{4}$. 
        }\label{FIG2}
  \end{center}
\end{figure}

The stationary limit of this correlation function reads:
\begin{widetext}
\beq\label{CAsymptotic}
\begin{array}{ll}
 \displaystyle 
\lim_{t\to\infty} \langle [\theta_A(x,t) - \theta_A(0,t) ]^2\rangle 
&  \displaystyle  =  \frac{1}{2\pi K}   \int  {\rm d}p \, (1 - \cos px)  e^{- \alpha^2 p^2} \Big[  \,  \frac{\pi}{p^2} \, \frac{1}{2 u} \sqrt{m_0^2+(u p)^2} +  \,  \frac{\pi u}{2} \frac{1}{\sqrt{m_0^2+(u p)^2} }  \Big] 
\\ \vspace{-0.2cm} \\
&  \displaystyle =  \frac{1}{K}   \int_0^{\infty}  {\rm d}p \, (1 - \cos px)  e^{- \alpha^2 p^2} \frac{1}{p} \cosh \Delta_p
\ ,
\end{array}
\eeq
\end{widetext}
and it turns out to be different from the thermal equilibrium correlation function.
Yet if $m_0 \alpha/u \gg 1$ one can expand and obtain:
\beq
\begin{array}{ll}
 \displaystyle 
\lim_{t\to\infty} \langle [\theta_A(x,t) - \theta_A(0,t) ]^2\rangle 
  \displaystyle  \simeq  \frac{1}{K}   \int_0^{\infty}  {\rm d}p \, (1 - \cos px)  \times
\\ \vspace{-0.2cm} \\
\displaystyle  \qquad\qquad\qquad \times e^{- \alpha^2 p^2} \, \frac{m_0}{2 u p^2}     \stackrel{x m_0/u \gg1}{\simeq}
 \frac{ \pi  m_0 x}{4 u K}
\ .
\end{array}
\eeq
This asymptotic form should be compared with the equilibrium: 
\beq\label{CThermal}
\begin{array}{ll}
\displaystyle \langle [\theta_A(x) - \theta_A(0) ]^2\rangle_{T_{\rm eff}}  = \displaystyle \frac{1}{K}   \int_0^{\infty}  {\rm d}p \, (1 - \cos px)  e^{- \alpha^2 p^2} \times
\\ \vspace{-0.2cm} \\
\qquad\qquad\qquad \displaystyle \times \frac{1}{p} \text{cotanh}\Big( \frac{u |p|}{2 T_{\rm eff} } \Big)
\\ \vspace{-0.2cm} \\
\qquad\qquad \displaystyle\simeq \displaystyle \frac{1}{K}   \int_0^{\infty}   {\rm d}p \, (1 - \cos px)  e^{- \alpha^2 p^2} \frac{2 T_{\rm eff} }{u p^2} 
\end{array}
\eeq
recovering $T_{\rm eff} = \frac{m_0}{4}$.
In the upper panel of Fig.~\ref{FIG2} we show the asymptotic correlation function $C_{\theta}(x,t\to\infty) = \exp(- \frac12 \lim_{t\to\infty} \langle [\theta_A(x,t) - \theta_A(0,t) ]^2\rangle) $
as given in (\ref{CAsymptotic}) with a black line and the thermal one with $T_{\rm eff} = \frac{m_0}{4}$ in blue. The two lines overlap remarkably 
and look indistinguishable.

\section{Dynamics of the density and current}\label{Sec_density}

\begin{figure}
  \begin{center}
        \includegraphics[height=5.5cm]{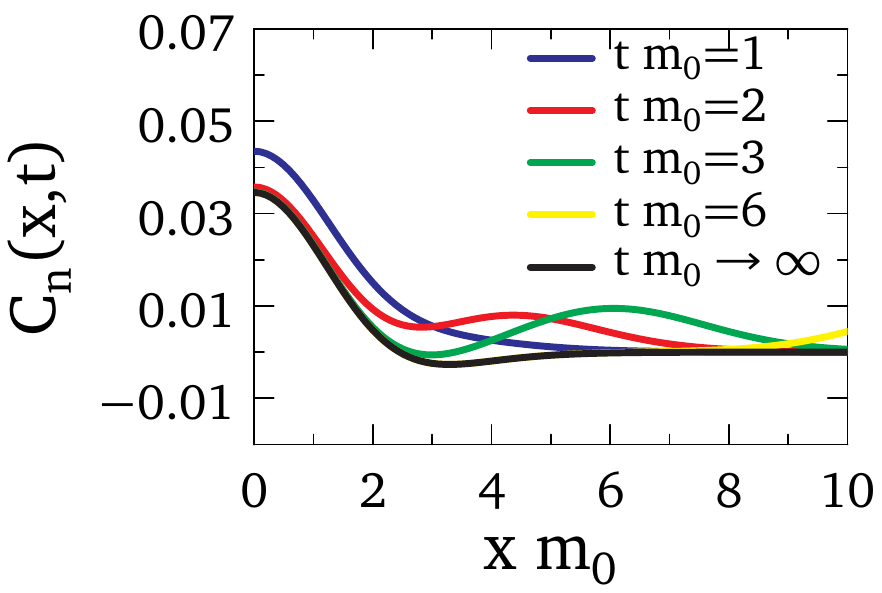}
    \includegraphics[height=5.5cm]{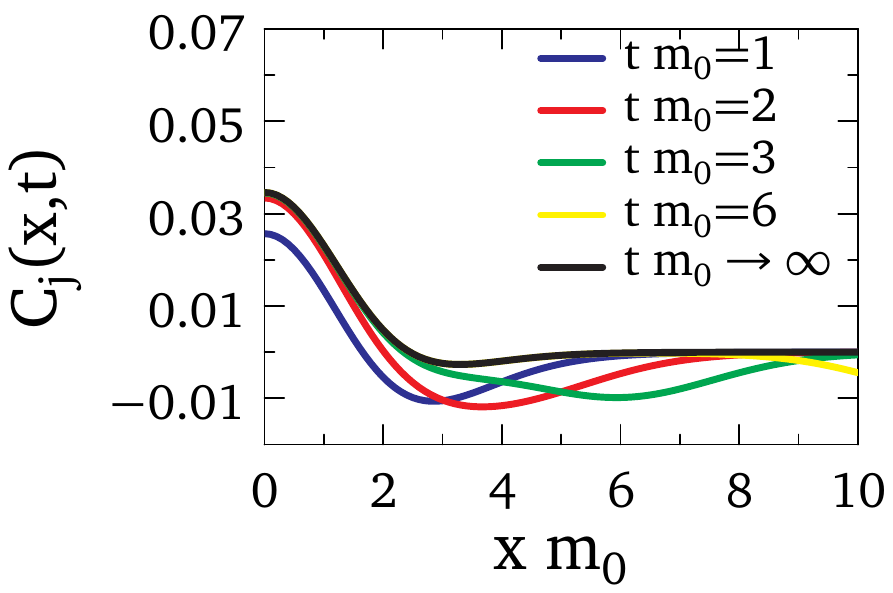}
        \caption{Correlation functions of the density $C_n(x,t)$ (upper panel) and 
        of the current $C_j(x,t)$ (lower panel) as a function of the space distance for different times.
        Times $t m_0 = 1,2,3,6$ are shown respectively with blue, red,  green and yellow lines. In both plots the black line is the
        asymptotic result in the stationary limit. In both figures we set $\alpha m_0 =1$, $K=1$ and $u=1$. 
        For the parameters used here the stationary curves in the two plots are the same.
        }\label{FIG3}
  \end{center}
\end{figure}

In this section we consider the dynamics of the relative density and current. 
The density can be experimentally recovered
by taking images of the two systems. 
We consider again the dynamics at $t_{\perp}=0$ which results in the following correlation function:
\beq\label{Cnxt}
\begin{array}{ll}
 \displaystyle C_n(x,t) = \langle n_A(x,t) n_A(0,t) \rangle 
= \frac{K}{4 \pi^2} \int \ {\rm d} p \ e^{i p x}  \ e^{-\alpha^2 p^2} \times
\\ \vspace{-0.2cm} \\
 \qquad\qquad\displaystyle \times \Big[  \sin^2 \omega_p t \, 
 \frac{ p^2  u}{ \omega_p^0}   + \,  \cos^2 \omega_p t \,  \frac{\omega_p^0}{ u}  \Big] \ .
  \end{array}
\eeq
The stationary value of this correlation function turns out to be:
\beq\label{Cnxtinfty}
\begin{array}{ll}
 \displaystyle C_n(x,t \to \infty) = \frac{K}{4 \pi^2} \int \ {\rm d} p \ e^{i p x}  \ e^{-\alpha^2 p^2} \, \frac12 \, \Big[   \frac{p^2  u}{\omega_p^0}   +   \frac{ \omega_p^0 }{u}\Big] 
\\ \vspace{-0.2cm} \\
\qquad\qquad \displaystyle =  \frac{K}{4 \pi^2} \int \ {\rm d} p \ e^{i p x}  \ e^{-\alpha^2 p^2} \, |p| \cosh \Delta_p
\ .
  \end{array}
\eeq
In the upper panel of Fig.~\ref{FIG3} we show the correlation function $C_n(x,t)$, as a function 
of the space distance for different times together with the asymptotic stationary result.
From the plot one sees that for sufficiently large times (e.g. $t m_0=2,3,6$) the dynamical curves follow 
the stationary limit up to a distance that grows with time   
and then departs from it.

The equilibrium correlation function is given by:
\beq\label{Cneq}
\begin{array}{ll}
 \displaystyle  \langle n_A(x) n_A(0) \rangle_{T_{\rm eff}} =&  \displaystyle  \frac{K}{4 \pi^2} \int \ {\rm d} p \ e^{i p x}  \ e^{-\alpha^2 p^2} | p| \times
\\ \vspace{-0.2cm} \\
  & \displaystyle \qquad\quad   \times\text{cotanh}\left( \frac{\omega_p}{2 T_{\rm eff}}  \right) \ .
 \end{array}
\eeq
In the lower panel of Fig.~\ref{FIG2} we show the asymptotic correlation function for the relative density~(\ref{Cnxtinfty}) with a black line
and the same correlation function at thermal equilibrium~(\ref{Cneq}) at $T_{\rm eff} = \frac{m_0}{4}$ with a blue line.
Also in this case the two functions are not the same but numerically they look indistinguishable
showing that for all practical purposes this state can be considered as a thermal ``equilibrium'' state.

It is natural to compare the correlation functions of the density with the ones involving the
current $j_A(x) =-\frac{1}{\pi} \nabla\theta_A(x) $. 
The corresponding correlation function reads:
\beq\label{Cjxt}
\begin{array}{ll}
 \displaystyle 
 \displaystyle C_j(x,t) =&  \displaystyle \langle j_A(x) j_A(0) \rangle 
 =  \displaystyle  \frac{1}{4 \pi^2 K} \int \ {\rm d} p \ e^{i p x}  \ e^{-\alpha^2 p^2} \times
\\ \vspace{-0.2cm} \\
&\displaystyle \qquad \times \Big[ \sin^2 \omega_p t \,    \frac{ \omega_p^0}{u} + \, \cos^2 \omega_p t \, \frac{u \, p^2 }{  \omega_p^0} \Big]

\end{array}
\eeq
The stationary limit of (\ref{Cjxt}) is the same as (\ref{Cnxtinfty}) with the substitution $K \to K^{-1}$.
The same substitution gives the equilibrium correlation function associated to the current from (\ref{Cneq}). 
In the lower panel of Fig.~\ref{FIG3} we show the correlation functions $C_j(x,t)$ as a function
of the space for different times, in particular $t m_0 = 1,2,3, 6$ are shown respectively with
blue, red,  green and yellow lines. The black line is the asymptotic result in the stationary limit.

\section{Discussions}\label{Sec_discussions}


Let us now discuss some aspects of the results obtained in the 
previous sections and the model we have used to obtain them. 

First we note that our results were obtained in the approximations of a quadratic expansion
of the cosine term that encodes the interaction between the two condensates 
around the minimum $\theta_A=0$. Such an approximation
is expected to work better in the weakly interacting regime (which is the one 
realized in~\cite{HLFSS2007,Gring_Science_2012,LGKRS13}) where the cosine term
is very relevant.

Under these assumptions
we have shown that the form of the initial state is
that of a squeezed state in terms of the operators that diagonalize
the final Hamiltonian at $t_{\perp}=0$.
This form is qualitatively the same as the one derived 
in~\cite{BA2007,KISD11,Gring_Science_2012} in order to describe
the state obtained from the splitting of a one dimensional gas into
two phase-coherent systems~\cite{HLFSS2007,Gring_Science_2012,LGKRS13}.
The coefficients appearing in the squeezed form are nonetheless different
and give raise to some differences in the physical quantities, as explained below.

Besides the characterization of the initial state,  
we have computed the correlation functions of the relative phase
between the two systems, correlations that are relevant for interference experiments.
From Fig.~\ref{FIG1} one sees that these correlations initially show long range order at large distances,
implying high interference contrast, and the value attained at large distances decays with time.
In comparison with the correlation functions computed in~\cite{LGKRS13} the qualitative form 
is similar with the appearance of a light cone that separates the length scales 
into a region of short distances growing with time 
where the correlation functions have relaxed and a region of larger distances that keeps track of the initial
correlation.
This light cone scenario has been observed also in other works~\cite{CC2006,IR2011,CEF2011,cheneau2012}
and it is justified with the creation at time $t=0$ of elementary excitations that propagate
in opposite direction and with finite velocity. In the case of $t_{\perp}=0$ the spectrum
is linear and all quasi-particles move coherently with the same group velocity and
two points at a given distance will equilibrate only when such quasi-particles 
have the time to travel along that length scale.

In addition to the correlation functions associated to the relative phase 
we have computed the evolution of the correlation functions
describing the relative density and current (see Fig.~\ref{FIG3}, upper and lower panel respectively).
These correlation functions do not show a clear light cone dynamics as for the
relative phase, however, in a similar way, at large times the dynamical curves stay 
close to the asymptotic stationary line up to distances that grow with time.

For all phase, density and current correlation functions we have considered  the limit of long times when they
all become stationary.
In this limit the analytical form of correlation functions is not compatible with that at thermal equilibrium, which 
is not surprising in view of the integrability of the dynamics and the Gaussian form
of the initial state here considered.
Nonetheless, numerically, stationary and thermal correlation functions are in remarkable agreement, 
so that experimentally one should not be able to distinguish between the two. The thermal correlation functions
that we have considered are associated with an effective temperature 
$T_{\rm eff} = \frac{m_0}{4}=  \frac14 \sqrt{\frac{J_0\rho u}{K} } $,
 where $m_0$ is the gap characterizing the spectrum before the quench, $J_0$ is the amplitude of the initial tunneling, $\rho$ the density,
 $u$ the speed of sound and $K$ le Luttinger parameter.
 
The agreement between the curves obtained as the stationary limit of the non-equilibrium process
and the ones drawn with the effective temperature  $T_{\rm eff} = \frac{m_0}{4}$ 
can be justified if one develops the asymptotic form of correlation functions at large $\frac{m_0 \alpha}{u}$, where 
$\alpha$ is the cut-off,
and develops the equilibrium 
correlation functions at large temperatures. However numerically one sees that the agreement is
good also at small values of $\frac{m_0 \alpha}{u}$.

Moreover we note that that if one assigns a different temperature to each mode by requiring the conservation
of their occupation numbers, as in (\ref{betaGGE}), the temperature $T_{\rm eff} = \frac{m_0}{4}$ is the one associated 
with the low energy modes.

Such an effective temperature, obtained within our quench protocol,  
should be contrasted with the one that it is
found with the initial state describing the splitting process $\tilde{T}_{\rm eff} =  \frac{\pi u \rho}{4 K} $~\cite{Gring_Science_2012}.
The two expressions clearly indicate a different dependence in the density. In particular,
in the weakly interacting regime where $\frac{u}{K}$ is independent of the density, 
the two procedures give raise respectively to a temperature that goes with $T_{\rm eff} \propto \sqrt{\rho}$
and one that is linear in the density $\tilde{T}_{\rm eff} \propto \rho$.

As we mentioned in the introduction, this quench protocol can be implemented experimentally, 
as ladders have already been realized~\cite{AALBPB2014}.
The same theoretical description can be applied to describe atom-chip experiments, 
once two one dimensional gases are equilibrated with a finite barrier which is subsequently
increased in order to suppress the tunneling between the two.
In both cases the effective temperature could be extracted from the decay of the
correlation functions  sufficiently long after the quench.

It would be therefore interesting to compare the results obtained for the effective temperature 
in the splitting process
with the ones of the quench that we propose.

\section{Conclusions}\label{Sec_conclusions}

In this work we have considered two chains that are tunnel coupled through
an interaction parameter $t_{\perp}$ which, at time $t=0$, is suddenly set to zero.
We have shown that the form of the state after the quench is that of a squeezed state,
similar (but not equal) to what was assumed in~\cite{KISD11,Gring_Science_2012} from phenomenological arguments. 

We have computed the correlation functions of the relative phase  which are directly probed in interference experiments 
and the correlation functions of the relative density and current.
For all of them we have extracted their asymptotic form as obtained at large times. 
We have shown that the stationary correlations are formally not compatible with the ones derived at thermal equilibrium.
Nonetheless numerically all correlation functions in the stationary state and those at thermal equilibrium
are in striking agreement. 
The effective temperature with which these correlation functions are compared is quite different from
the one derived in~\cite{KISD11,Gring_Science_2012,LGKRS13}. In our quench it is associated with 
the mass of the  initial Hamiltonian and depends on the squared root of the density 
while the effective temperature found in ~\cite{KISD11,Gring_Science_2012,LGKRS13} 
is linear in the density. 

In view of this difference it would be interesting to compare the effective temperature 
that is found after the splitting of the condensate with the one that one recovers considering
the quench suggested in this work.

As a future work, it would be interesting to study the effect of  the initial temperature on the subsequent
dynamics, by considering as initial state a density matrix instead of the ground state of the
Hamiltonian for some $t_{\perp}$. In fact, in~\cite{KISD11,Gring_Science_2012,LGKRS13} it was found that the effective temperature
is insensitive to the temperature of the system before the splitting and it would be 
worth investigating up to which extent this property carries over in our quench protocol.

\acknowledgments 

We thank J. Schmiedmayer for many interesting discussions. 
This work was supported in part by the Swiss National Science Foundation under Division II 
and also by the ARO-MURI Non-equilibrium Many-body Dynamics grant (W911NF-14-1-0003).

\bibliographystyle{mioaps}
\bibliography{dynamics}

\end{document}